\documentclass[a4paper,noshowpacs,noshowkeys,prb,12pt]{revtex4}

\usepackage{amsmath}
\usepackage{amssymb}

\providecommand{\eq}[1]{\begin{equation} #1 \end{equation}}

\providecommand{\aver}[1]{\langle #1 \rangle}
\providecommand{\mt}[1]{\mbox{\tiny $#1$}}
\providecommand{\ms}[1]{\mbox{\small $#1$}}
\providecommand{\mn}[1]{\mbox{\normalsize $#1$}}
\providecommand{\ml}[1]{\mbox{\large $#1$}}
\providecommand{\bs}[1]{\boldsymbol{#1}}

\providecommand{\Tr}{{\rm Tr}}

\font\bb=bbmss12 scaled 1000
\def\id{\mbox{\bb 1}}

\begin{document}

\title{\Large Simple derivation of general Fierz-type identities}

\author{\large C. C. Nishi}
\email{ccnishi@ift.unesp.br}
\affiliation{Instituto de F\'\i sica Te\'orica,
Universidade Estadual Paulista,
Rua Pamplona 145, 01405-900, S\~ao Paulo, SP, Brazil}

\begin{abstract}
\setlength{\baselineskip}{.68\baselineskip}
\large
General Fierz-type identities are examined and their well known
connection with completeness relations in matrix vector spaces is
shown. In particular, I derive the chiral Fierz identities in a
simple and systematic way by using a chiral basis for the complex
$4\times4$ matrices. Other completeness relations for the
fundamental representations of $SU(N)$ algebras can be extracted
using the same reasoning.
\end{abstract}




\maketitle

\section{Introduction}
The Fierz identities\cite{fierz} are frequently used in particle physics to
analyze four-fermion operators\cite{DGH} such as current-current operators.
These reordering identities are used to write a product of two Dirac
bilinears\cite{bjorken} as a linear combination of other products
of bilinears with the four constituent Dirac spinors in a different
order. Because Fierz identities relate Dirac bilinears, which are
objects with well defined transformation properties under the
Lorentz group, and Dirac gamma matrices form a representation of
the Lorentz group generators, it is not surprising that Fierz
identities imply the existence of a basis of four-vectors formed by
Dirac bilinears and guarantee its orthogonality and completeness
properties.\cite{takahashi}
The converse is also possible, that is, recovering the spinor from that basis,
which reveals the equivalence between spinor and tensorial representation of
various quantities.

Fierz identities are only a particular set of matrix identities,
valid for the Dirac bilinears which span the space of $4\times 4$
complex matrices. General Fierz-type identities can be found for
any $N\times N$ real or complex square matrices.
The primary aim of this article is to show that any Fierz-type
relation can be deduced using just a few ingredients such as the
notion of a basis in a vector space and its properties of
orthogonality and completeness. The relevant vector space in this
case is the vector space of square matrices. For such a vector
space, a basis and the orthogonality property between its elements
can be defined through an inner product. These ideas are explained
in detail in Sec.~\ref{sec1} for general square matrix vector
spaces. Some particularly useful examples involving Pauli matrices
(SU(2) algebra), Gell-Mann matrices (SU(3) algebra), and
fundamental representations of general SU(N) algebras are given.

As a particular nontrivial application of these ideas, we will find
a straightforward way to deduce the chiral Fierz identities, that
is, Fierz identities involving bilinears containing left/right
chiral projectors. One way to obtain the chiral Fierz identities
is to use the original Fierz identities; one can find very general Fierz
identities for usual bilinear  in Ref.\,\onlinecite{takahashi}.
However, this procedure may become quite lengthy because it requires expanding
the left/right projected bilinears in terms of the usual bilinears, performing
Fierz reorderings, and then rewriting them as projected bilinears.
Such an approach was adopted in Ref.\,\onlinecite{nieves}, but will not be
pursued here. Instead, the chiral Fierz identities will be deduced by rederiving
the Fierz identities using appropriate left/right projected bilinears as a basis
(Sec.~\ref{sec3}). The simple form of certain chiral Fierz identities is already
an indication that a much simpler procedure should exist to derive
them.

To exemplify the practical importance of chiral Fierz identities, I
will show a situation where the reordering permitted by chiral
Fierz identities can help us to better understand and more simply
describe a physical system. The obvious area where the chiral Fierz
identities can be important is the physics involving the weak
interaction, which is a mainly left-handed interaction. The
particular physical system concerns
the neutrinos propagating through ordinary matter. Because ordinary
matter contains electrons but is absent of muons and tauons, only
the electron neutrinos interact with the electrons in matter
through the charged-current effective interaction
Lagrangian\cite{Mohapatra}
\eq{
\label{weak:1}
\frac{4G_F}{\sqrt{2}} \bar{\psi}_e(x)\gamma^{\alpha}L\psi_{\nu_e}(x)
\bar{\psi}_{\nu_e}(x)\gamma_{\alpha}L\psi_{e}(x),
}
where $\psi_e,\psi_{\nu_e}$ are the fields for the electron and the
electron neutrino respectively, and $L=(\id-\gamma_5)/2$ is the projector for
left chirality. Using a chiral Fierz identity we can rewrite
Eq.~\eqref{weak:1} in the form
\eq{
\label{weak:2}
\frac{4G_F}{\sqrt{2}}
\bar{\psi}_{\nu_e}(x)\gamma^{\alpha}L\psi_{\nu_e}(x)\,
\bar{\psi}_e(x)\gamma_{\alpha}L\psi_{e}(x).
}
This reordered form enables us to describe the influence of
non-relativistic electrons in matter by their average density
$2\aver{\bar{\psi}_e(x)\gamma_{\alpha}L\psi_{e}(x)}\sim \delta_{\alpha
0}n_e$. Such a term leads to an effective interaction acting on electron
neutrinos that is different from the interactions acting on other types of
neutrinos. Ultimately, it leads to a significant modification of the
description of neutrino oscillations,\cite{Mohapatra} the phenomenon
responsible for the missing solar neutrino problem.\cite{neutrinos}

Before we get into the details, let us consider the difference between the
expressions in Eqs.\,\eqref{weak:1} and \eqref{weak:2} to understand better  the
nonintuitive nature of the Fierz identity.
The spinor fields $\psi_{e}$ and $\psi_{\nu_e}$ can be written as complex
 $4\times 1$
matrices, while $\bar{\psi}_{e}$ and $\bar{\psi}_{\nu_e}$ are complex $1\times
 4$
matrices. The factors between them are $4\times 4$ matrices and the result
is a scalar. In Eq.\,\eqref{weak:1}, there is a
$4\times 4$ matrix between $\bar{\psi}_e$ and $\psi_{\nu_e}$ and another between
$\bar{\psi}_{\nu_e}$ and $\psi_{e}$ for each $\alpha=0,1,2,3$. What the Fierz
identities assure you is that any expression of the form
$\bar{\psi}_1A\psi_{2}\bar{\psi}_{3}B\psi_{4}$ is equal to
$\bar{\psi}_1C\psi_{4}\bar{\psi}_{3}D\psi_{2}$ or a sum of similar
terms, with $A,B\neq C,D$ in general. Amazingly, the same combination of
matrices that enters into Eq.\eqref{weak:1} also enters into Eq.\eqref{weak:2}
as a consequence of a Fierz identity.

\section{Matrix Vector Spaces and Completeness Relations}
\label{sec1}

A vector space $V$ is a set of elements endowed with two
operations:\cite{LA} a sum between elements and a
multiplication by numbers (elements of a {\em ring}, in
mathematical language) such as the real and complex numbers. This vector
space is required to be closed under such
operations.\cite{endnote1}
The usual $N$-dimensional real space, $\mathbb{R}^N$,
and its complex extension, $\mathbb{C}^N$, are examples of vector spaces.

As is well known, any element of a vector space can be expanded in
terms of a basis $\{e_i\}$, a set of $N={\rm dim}V$ elements.
In addition, the vector space can be equipped with an inner product
$(~,~)$ that defines the notion of norm and orthogonality.
By using such an inner product, an orthonormal basis $\{e_i\}$ can be defined
by
\eq{
\label{ortho1}
(e_i,e_j)=\delta_{ij}.
}
The canonical (column vector) representation for the orthonormal basis
is
\eq{
\label{canon}
(e_i)_j=\delta_{ij},
}
where the index $j$ outside the parenthesis is the vector index.
In matrix notation the orthogonality relation \eqref{ortho1} can be
written as
\eq{
e_i^{\mt{\top}}e_j=\delta_{ij}.
}
where ${\mt{\top}}$ denotes the transpose. In canonical form, the
completeness relation
\eq{
\label{complete1}
\sum_{i}^{N}e_i\,e_i^{\mt{\top}}=\id,
}
is obvious. Also, Eq.~\eqref{complete1} is
invariant under an orthogonal $O(N)$ (unitary $U(N)$) transformation of
a basis for $V=\mathbb{R}^N (\mathbb{C}^N)$.

Once the properties of vector spaces are given, it is easy to show
that the set of all $N\times N$ square matrices over the
reals, $M_{N}(\mathbb{R})$, or over the complex numbers,
$M_N(\mathbb{C})$, form a $N^2$ dimensional vector
space.\cite{endnote2}
In these vector spaces a canonical basis $\{e^{ij}\}$ is given by
the matrices
\eq{
\label{basis}
(e^{ij})_{kl}=\delta^i_k\,\delta^j_l \quad
(\ms{i,j,k,l=1,\cdots,N}), }
and hence, any $N\times N$ matrix can be expanded as
\eq{
\label{expansion}
M=M_{ij}\,e^{ij},
}
where the expansion coefficients $M_{ij}=(M)_{ij}$ are the elements
of the matrix $M$.

In $M_N(\mathbb{R})$ we can define the (positive definite) inner
product \eq{
\label{inner}
(A,B)\equiv \Tr[AB^{\mt{\top}}],
}
for which the canonical basis satisfies
\eq{
\label{ortho2}
\Tr[e^{ij}{e^{kl}}^{\mt{\top}}]=\delta^{ik}\delta^{jl}.
}
For $M_N(\mathbb{C})$ the transpose operation $(~)^{\top}$ has to be
replaced by the hermitian conjugation operation $(~)^\dag$.

An equivalent approach is to define a bilinear function on
$M_N(\mathbb{R})$ without the positive definiteness requirement
of an inner product. Thus,
instead of defining Eq.~\eqref{inner}, we can discard the transpose
operation in the definition and regard simply the trace of the
product as the relevant bilinear function. The missing transpose
operation can be transferred to the definition of a dual basis
$\{e_{ij}\}$:
$e_{ij}\equiv e^{ji}={e^{ij}}^{\mt{\top}}$. Then the dual basis is
the orthogonal counterpart of the basis $\{e^{ij}\}$ through the
trace bilinear. An equivalent way is to regard the trace bilinear
between two elements of $\{e^{ij}\}$ as defining a metric that can
be used to lower and raise indices and to interchange the
basis with its dual; an analogous construction is found in special
relativity when contravariant and covariant four-vectors are
defined. An inner product defines, with an appropriate basis, a
metric that is the identity matrix. In general there can
be non-diagonal or non-positive definite metrics. This
approach will be used to derive the Fierz identities in
Sec.~\ref{sec2} and chiral Fierz identities in Sec.~\ref{sec3}.

We use this dual basis to express the expansion
coefficient in Eq.~\eqref{expansion} as
\eq{
\label{coef}
M_{ij}=\Tr[Me_{ij}].
}
If we substitute Eq.~\eqref{coef} in Eq.~\eqref{expansion} and take
the
respective elements of the matrix, we obtain the trivial
relation
\eq{
\label{complete2}
\delta_{km}\delta_{nl}=
(e_{ij})_{kl}(e^{ij})_{nm},
}
where the summation convention of repeated indices is used here and
in the following. This relation follows directly from
Eq.~\eqref{canon} and is a completeness relation analogous to
Eq.~\eqref{complete1}. Equation \eqref{complete2} represents an
identity in
the space of general linear transformations over
$M_N(\mathbb{R})$. This entire discussion is also valid for
$M_N(\mathbb{C})$ if one extends the ring from $\mathbb{R}$ to
$\mathbb{C}$.
Because any linear transformation over $M_N(\mathbb{R})$ can be
given as a linear combination of transformations of the form
\eq{
M \rightarrow
(A\otimes B)M \equiv AMB^{\top} =(A)_{ij}(B)_{lk}(M)_{jk}\,e^{il},
}
Eq.~\eqref{complete2} implies that
\eq{
(e_{ij}\otimes e^{ij})M=(e^{ij}\otimes e_{ij})M
=M.
}
Although trivial with this choice of basis, a completeness relation like
Eq.~\eqref{complete2} is all that is needed to deduce Fierz-type
identities.

To obtain nontrivial relations, let us take the example of
SU(2) and SU(3) algebras in the fundamental representation. The
commonly used representations for these algebras are the Pauli
matrices
$\{\sigma_i\}$ and the Gell-Mann matrices
$\{\lambda_a\}$.\cite{IZ,endnote3} They form vector spaces and satisfy
the orthogonality relations
\begin{align}
\Tr[\sigma_i\sigma_j]&=2\delta_{ij} \quad (i,j=1,2,3)\\
\Tr[\lambda_a\lambda_b]&=2\delta_{ab} \quad (a,b=1,\ldots,8).
\end{align}
Because they are already orthogonal and are hermitian matrices,
there is no need to define a dual basis.
However, to span $M_2(\mathbb{C})$ and $M_3(\mathbb{C})$ the
respective identity matrices are needed, because
$N^2$ basis vectors are required and the Pauli and Gell-Mann
matrices are traceless. Then the set $\{\id,\sigma_i\}$
spans $M_2(\mathbb{C})$, which means any $2\times 2$
complex matrix can be expanded in terms of
\eq{ \label{pauli}
X=X_0\id+X_i \sigma^i, ~~\sigma^i=\sigma_i,
}
where
\eq{
\label{above}
X_0=\frac{1}{2}\Tr[X], \quad X_i=\frac{1}{2}\Tr[X\sigma_i].
}
We substitute Eq.~\eqref{above} into Eq.\eqref{pauli} and take the
general elements
\eq{
(X)_{ij}=\frac{1}{2}(X)_{kk}\delta_{ij}
+\frac{1}{2}(X)_{lk}(\sigma_m)_{kl}(\sigma_m)_{ij},
}
and obtain from the coefficients of $(X)_{lk}$, after properly
inserting Kronecker deltas, the completeness relation
\eq{
\label{complete3}
\delta_{il}\delta_{kj}=\frac{1}{2}\delta_{ij}\delta_{kl}
+ \frac{1}{2}(\sigma_m)_{ij}(\sigma_m)_{kl}.
}
Equation~\eqref{complete3} is the identity used to deduce the Fierz
identities to Weyl spinors:
$(\sigma_{\mu})_{ij}(\tilde{\sigma}^{\mu})_{kl}=2\delta_{il}\delta_{kj}$,
where $\sigma^{\mu}=(\id,\bs{\sigma})$, and
$\tilde{\sigma}^{\mu}=(\id,-\bs{\sigma})$; the lowering and
raising of $\mu$ indices follows the Minkowski metric.

For the Gell-Mann matrices we have similarly,
\eq{
\label{gellmann}
\frac{1}{2}(\lambda_a)_{ij}(\lambda_a)_{kl}+\frac{1}{3}\delta_{ij}
\delta_{kl} =\delta_{il}\delta_{kj}.
}
For any fundamental representation of SU(N) algebra $\{T_a\}$
satisfying
$
\Tr[T_aT_b]=C\delta_{ab},
$
we have the completeness relation
\eq{
\label{sun}
\frac{1}{C}(T_a)_{ij}(T_a)_{kl}+\frac{1}{N}\delta_{ij}
\delta_{kl} =\delta_{il}\delta_{kj}.
}

For the $O(N)$ groups there is no simple relation similar to
Eq.~\eqref{sun} because the algebra is formed by $N\times N$
antisymmetric matrices, and thus all symmetric matrices are needed
to span $M_N(\mathbb{R})$.

Before introducing Dirac matrices to deduce the Fierz
identities, it is better to introduce a clearer notation due to
Takahashi,\cite{takahashi} where we replace the matrix indices
by parentheses (~) and brackets [~\,], such that each
parenthesis/bracket represents a different index in an unambiguous way.
For example, using this notation the relation \eqref{sun} reads
\eq{
\label{completeN}
\frac{1}{C}(T_a)[T_a]+\frac{1}{N}(~)[\,~]=
(\,~][~\,),
}
where the blank entry means the identity matrix.
This notation clearly shows the reordering property.

\section{Fierz Identities}
\label{sec2}

The starting point to derive the usual Fierz identities is the
orthogonality relation among the 16 Dirac bilinears\cite{endnote4}
that span $M_4(\mathbb{C})$ over $\mathbb{C}$. The 16 Dirac
bilinears are usually classified into distinct classes according to
their properties under Lorentz transformations\cite{bjorken} as
\eq{
\label{gamma1}
\{\Gamma^A\}=
\{\id,\gamma_5,\gamma^{\mu},\gamma_5\gamma^{\mu},\sigma^{\mu\nu}\}
\quad (\mu,\nu=0,1,2,3), }
where $\mu<\nu$ in $\sigma^{\mu\nu}$ to avoid redundancy.
Here the convention used by Itzykson and Zuber\cite{IZa}
is employed for the gamma matrices:
\begin{align}
\{\gamma^{\mu},\gamma^{\nu}\}&=2g^{\mu\nu}\\
\sigma^{\mu\nu}&=\frac{i}{2}[\gamma^{\mu},\gamma^{\nu}]\\
\gamma^5& =\gamma_{5}=i\gamma^0\gamma^1\gamma^2\gamma^3\\
\gamma^5\sigma^{\mu\nu}&=\frac{i}{2}\varepsilon^{\mu\nu\alpha\beta}
\sigma_{\alpha\beta},
\end{align}
where $\varepsilon^{0123}=1= -\varepsilon_{0123}$ and $g^{\mu\nu}={\rm
diag}(1,-1,-1,-1)$ is the Minkowski metric.

Then, defining a basis $\{\Gamma_A\}$ dual to Eq.~\eqref{gamma1}
as the respective gamma matrices with space-time indices lowered by
Minkowski metric,\cite{endnote5} the orthogonality relation holds:
\eq{
\label{inner1}
\Tr[\Gamma_A\Gamma^B]=4\delta_A^B~.
}
This relation allow us to expand any complex $4\times 4$ matrix $X$ in
terms of the basis \eqref{gamma1} as
\eq{
\label{expansion2}
X=X_A\Gamma^A, \quad
X_A=\frac{1}{4}\Tr[X\Gamma_A].
}
We combine Eqs.~\eqref{inner1} and \eqref{expansion2}, extract each element
of the matrix, and find a completeness relation analogous to
Eq.~\eqref{completeN}:
\eq{
\label{completeD}
(~)[\,~]=\frac{1}{4}(\Gamma_A][\Gamma^A)=\frac{1}{4}(\Gamma^A][\Gamma_
A ) .
}
This identity is sufficient to reproduce all possible Fierz identities by
appropriately multiplying identity matrices by general
matrices $X,Y$ as
\begin{align}
\label{fierz1}
(X)[Y]&=(X\id)[\id Y]=\frac{1}{4}(X\Gamma_CY][\Gamma^C)
\nonumber \\
&=\frac{1}{4^2}\Tr[
X\Gamma_CY\Gamma_D](\Gamma^D][\Gamma^C).
\end{align}
In particular, if $X=\Gamma^A$ and $Y=\Gamma^B$, Eq.~\eqref{fierz1}
leads to the Fierz identities
\eq{
\label{fierz2}
(\Gamma^A)[\Gamma^B]=
\frac{1}{4^2}\Tr[
\Gamma^A\Gamma_C\Gamma^B\Gamma_D](\Gamma^D][\Gamma^C).
}
The only remaining task is to calculate the expansion coefficients which
are straightforward gamma matrix traces.\cite{IZa}
The usual textbook Fierz identities (see, for example, Ref.~\onlinecite{IZ},
p.~160) can be found when $\Gamma^A$ and $\Gamma^B$ in Eq.~\eqref{fierz2}
are chosen to form scalar quantities (under the full Lorentz transformations,
including parity) such as $(\gamma_{\mu})[\gamma^{\mu}]$ or
Eq.~\eqref{completeD} itself. An additional
minus sign arises in the Fierz identities \eqref{fierz2} when we insert
anticommuting fermion fields instead of numerical spinors.

\section{Chiral Fierz Identities}
\label{sec3}

The Fierz identities derived in Sec.~III are not quite
appropriate when treating chirally projected combinations such as
\eq{
\label{example1}
(R\gamma^{\mu})[L\gamma_{\mu}],
}
where the two chiral projectors are $R=\frac{1}{2}(\id+\gamma_5)$ and
$L=\frac{1}{2}(\id-\gamma_5)$ because the expansion \eqref{fierz1} applied
to Eq.~\eqref{example1} still have some non-null coefficients to be
calculated. For non-scalar combinations such as
$(R\sigma^{\mu\nu})[R\gamma_{\nu}]$, the number of coefficients to be
calculated may become large. Moreover, the relatively simple form of certain
chiral Fierz identities such as the form invariants\cite{endnote6}
\begin{subequations}
\begin{align}
(R\gamma^{\mu})[R\gamma_{\mu}]
&=-(R\gamma^{\mu}][R\gamma_{\mu}) \\
(L\gamma^{\mu})[L\gamma_{\mu}]
&=-(L\gamma^{\mu}][L\gamma_{\mu}),
\end{align}
\end{subequations}
suggests a simpler procedure should exist.

A better way to perform Fierz transformations
for combinations such as Eq.~\eqref{example1} consists of
rederiving Fierz identities using a chiral basis
\eq{
\label{chiralbasis}
\{\Gamma^A\}=
\{R,L,R\gamma^{\mu},L\gamma^{\mu},\sigma^{\mu\nu}\} \quad
(\mu,\nu=0,1,2,3), }
where $\mu<\nu$, and its respective dual basis
\eq{
\label{chiralbasis2}
\{\Gamma_A\}=
\{R,L,L\gamma_{\mu},R\gamma_{\mu},\frac{1}{2}\sigma_{\mu\nu}\}
\quad (\mu,\nu=0,1,2,3), }
where $\mu<\nu$.
Notice that because of the anticommuting nature of $\gamma^{\mu}$ with
$\gamma_5$ and projector properties, the dual of $R\gamma^{\mu}$ is
$L\gamma_{\mu}$. The orthogonality property between the
bases \eqref{chiralbasis} and \eqref{chiralbasis2} is
\eq{
\label{ortho3}
\Tr[\Gamma_A\Gamma^B]=2\delta_A^B,
}
which implies the completeness relation
\begin{align}
\label{completeC}
\!(~)[\,~] & = \ml{\frac{1}{2}}(\Gamma_A][\Gamma^A)
\nonumber \\
&= \ml{\frac{1}{2}}\{
(R][R)+(L][L)+(R\gamma^{\mu}][L\gamma_{\mu}),
+\,(L\gamma^{\mu}][R\gamma_{\mu})
+(\ms{\frac{1}{2}}\sigma^{\mu\nu}][\ms{\frac{1}{2}}\sigma_{\mu\nu})
\},
\end{align}
where the extra $\frac{1}{2}$ in the $\sigma^{\mu\nu}$ expansion is
inserted to account for the double counting due to the implicit
$\mu,\nu$ summation over all values and
$\sigma^{\mu\nu}=-\sigma^{\nu\mu}$. Such a completeness relation directly
leads to the chiral Fierz identities
\eq{
\label{CFI}
(\Gamma^A)
[\Gamma^B]
=\frac{1}{4}\Tr[\Gamma^A\Gamma_C\Gamma^B\Gamma_D]
(\Gamma^D][\Gamma^C).
}

To illustrate the usefulness of the chiral Fierz identities, we
apply them to calculate the Fierz transform of the combination
\eqref{example1},
\begin{align}
(R\gamma^{\mu})[L\gamma_{\mu}]
&=\ml{\frac{1}{4}}\Tr[R\gamma^{\mu}\Gamma_CL\gamma_{\mu}\Gamma_D]
(\Gamma^D][\Gamma^C)
\nonumber \\
&=
\raisebox{1.1em}{}{}
\ml{\frac{1}{4}}\Tr[R\gamma^{\mu}LL\gamma_{\mu}R]
(R][L) =
2(R][L),
\end{align}
where the gamma matrix properties\cite{IZa}
$\gamma^{\mu}\gamma_{\mu}=4\!\times\!\id$,
$\gamma^{\mu}\sigma^{\alpha\beta}\gamma_{\mu}=0$,
and the trace cyclic property were used.
More difficult examples can be worked out, for instance,
\begin{align}
(R\sigma_{\mu\nu})[R\gamma^{\nu}]
& =
\mn{\frac{1}{4}}\Tr[R\sigma_{\mu\nu}RR\gamma^{\nu}L\gamma_{\rho}]
(R\gamma^{\rho}][R)
+
\mn{\frac{1}{4}}\Tr[R\sigma_{\mu\nu}\ms{\frac{1}{2}}\sigma_{\alpha\beta}
R\gamma^{\nu}L\gamma_{\rho}]
(R\gamma^{\rho}][\ms{\frac{1}{2}}\sigma^{\alpha\beta})
\nonumber \\
&=
\mn{\frac{3}{2}}i\,(R\gamma_{\mu}][R)
+\mn{\frac{1}{2}}(R\gamma^{\nu}][R\sigma_{\nu\mu}).
\end{align}
Some labor can be saved by using the trace relation
\eq{
\Tr [\underset{\mn{L}}R\sigma_{\mu\nu}\sigma_{\alpha\beta}]=
2(g_{\mu\alpha}g_{\nu\beta}-g_{\mu\beta}g_{\nu\alpha}
\pm i\epsilon_{\mu\nu\alpha\beta})
~.
}
One can check the coeficients for the cases $(\mu\nu)=(\alpha\beta)$ and
$(\mu\nu\alpha\beta)=(0123)$.

Moreover, the combination of chiral Fierz identities \eqref{CFI} with other
completeness relations such as Eq.~\eqref{gellmann} can be used to decompose
four-quark operators carrying other quantum numbers like
$SU(3)$ color.\cite{DGH}

The simplicity arises because only a few expansion coefficients are
nonzero due to the projector properties of $R/L$ and the commuting
or anticommuting character of the bilinears with $\gamma_5$.
Equivalently,
$R/L$ projectors reduce the spinor vector space and the resulting
projected spinors have to be the same on the two sides of the
chiral Fierz identities.

\section{Summary}
\label{sec4}

The well known result that the Fierz identities are
a consequence of the completeness of the Dirac bilinears as a basis
spanning the complex
$4\times 4$ complex matrices was reviewed.
Recognizing that bases other than Dirac bilinears are
equally possible permitted us to develop a better procedure
to calculate the chiral Fierz identities by choosing chirally
left/right projected matrices as a basis. The generality of the
procedure was stressed and
illustrated using the canonical basis of matrix vector spaces,
which led to trivial relations in this case.

The usefulness of Fierz-type relations depends on the particular
choice of basis and how nearly complete is the set of matrix objects
(representations) of interest.
The same unified framework was used to derive completeness relations
for the generators of the SU(N) group in the fundamental
representation. Other matrix representations or
other algebras can be analyzed as well, although they may not be
complete and hence the corresponding Fierz-type identities may not
be as useful as those presented here.

\acknowledgments

This work was supported by Conselho Nacional de Desenvolvimento
Cient\'\i fico e Tecnol\'ogico (CNPq).



\begin{thebibliography}{99}

\bibitem{fierz} M.~Fierz, ``Zur Fermischen Theorie des $\beta$-Zerfalls,''
Z. Physik {\bf 104}, 553-565 (1937).

\bibitem{DGH} J. F. Donoghue, E. Golowich, and B. R. Holstein, {\em
Dynamics of the Standard Model} (Cambridge University Press, 1994),
pp. 217 and 221.

\bibitem{bjorken} J. D. Bjorken and S. D. Drell, {\em Relativistic
Quantum Mechanics} (McGraw-Hill, 1965).

\bibitem{takahashi} Y. Takahashi, ``The Fierz identities,'' in {\em
Progress in Quantum Field Theory}, edited by H. Ezawa and S.
Kamefuchi (North-Holland, 1986), p. 121.

\bibitem{nieves} J. F. Nieves and P. B. Pal, ``Generalized Fierz
identities,'' Am. J. Phys. {\bf 72}, 1100--1108 (2004).

\bibitem{Mohapatra} R. Mohapatra and P. Pal, {\em Massive Neutrinos
in Physics and Astrophysics} (World Scientific, Singapore, 1991),
pp. 165--70.

\bibitem{neutrinos} The surprising implications and experimental
evidences for neutrino oscillations can be found in W. C. Haxton
and B. R. Holstein, ``Neutrino physics,'' Am. J. Phys. {\bf 68},
15--32 (2000); also ibid., ``Neutrino physics: An update,'' Am. J.
Phys. {\bf 72}, 18--24 (2004). See also M. C. Gonzalez-Garcia and
Y. Nir, ``Neutrino masses and mixing: Evidence and implications,''
Rev. Mod. Phys. {\bf 75}, 345--402 (2003).

\bibitem{LA} See any linear algebra book, for example, I. M. Gel'fand,
{\em Lectures on Linear Algebra} (Interscience, 1961).

\bibitem{endnote1} Any operation applied to an element
in the vector space must result in another element of the vector space.

\bibitem{endnote2} $M_N(\mathbb{C})$ may be considered as a
$2N ^2$ dimensional vector space if spanned by $N^2$ real matrices and
$N^2$ purely complex matrices with real expansion coefficients only, that
is, over the reals $\mathbb{R}$.

\bibitem{IZ} See C. Itzykson and J. B. Zuber, {\em Quantum Field Theory}
(McGraw-Hill, 1980), p. 516,
for an explicit representation of Gell-Mann matrices.

\bibitem{endnote3} The SU(2) and SU(3) generators are
\{$\frac{1}{2}\sigma_i$\} and $\{\frac{1}{2}\lambda_a\}$, respectively.

\bibitem{endnote4} I will denote as Dirac bilinears the proper
bilinears containing the two spinors as well as the associated matrices
alone, because the Fierz identities do not depend on the spinors involved.

\bibitem{endnote5} The lowering of space-time indices is equivalent
to the hermitian conjugation operation.

\bibitem{IZa} See Ref.~\onlinecite{IZ}, appendix.

\bibitem{endnote6} These identities were used to get from
Eq.~\eqref{weak:1} to Eq.~\eqref{weak:2} with a sign difference due to the
anticommutation of fermion fields.
\end{thebibliography}
\end{document}